\documentclass[aps,twocolumn,showpacs,superscriptaddress]{revtex4-1}
\usepackage{amsfonts}
\usepackage{amsmath}
\usepackage{amssymb}
\usepackage{lipsum}
\usepackage{graphicx}
\usepackage{bm}
\usepackage{natbib}
\usepackage{float}
\usepackage{multirow}
\usepackage{ulem}
\usepackage[dvipsnames]{xcolor}

\def\iu{\mathrm{i}}

\begin{document}
\title{Generalized dynamical mean-field theory of two-subalttice systems with long range interactions and its application to study charge and spin correlations in graphene}
\author{A. A. Katanin}
\affiliation{Center for Photonics and 2D Materials, Moscow Institute of Physics and Technology, Institutsky lane 9, Dolgoprudny, 141700, Moscow region, Russia}
\affiliation{M. N. Mikheev Institute of Metal Physics of Ural Branch of Russian Academy of Sciences, S. Kovalevskaya Street 18, 620990 Yekaterinburg, Russia}
\date{\today}
\begin{abstract}
We investigate magnetic and charge correlations in graphene by using the 
formulation of extended dynamical mean-field theory (E-DMFT) for two-sublattice systems.
First, we map the average non-local interaction onto the effective static interaction between different sublattices, which is treated together with the local interaction within an effective ``two-orbital'' local model. The remaining part of the non-local interaction is considered by introducing an effective retarded interaction within the E-DMFT approach. The non-local susceptibilities in charge and spin channel are further evaluated in the ladder approximation. We verify the applicability of the proposed method to describe the effect of uniformly screened long-range Coulomb potential $\propto 1/r$, as well as screened realistic long-range electron interaction [T. O. Wehling et. al., Phys. Rev. Lett. {\bf 106}, 236805 (2011)] in graphene. We show that the developed approach describes a  competition of semimetal,
spin density wave (SDW), 
and charge-density-wave (CDW) 
correlations. The obtained phase diagram is in a good agreement with recent results of functional renormalization group (fRG) for finite large graphene nanoflakes
and scaling analysis of quantum Monte Carlo data on finite clusters.  
Similarly to the previously obtained results within the fRG approach, the realistic screening of  Coulomb interaction by $\sigma$ bands causes moderate (strong) enhancement of critical long-range interaction strength, needed for the SDW (CDW) instability, compared to the results for the uniformly screened Coulomb potential. 
\end{abstract}
\maketitle
\section{Introduction}
Graphene, which is a two-dimensional allotrope form of carbon, being discovered experimentally in 2005 \cite{Graphene}, still attracts attention of both experimental and theoretical research due to its unusual properties. In particular, this substance is characterized by fermionic excitations with Dirac form of electronic dispersion, which is  similar to that for relativistic massless particles, albeit with the velocity 300 times smaller than speed of light. Dirac form of electronic dispersion yields in particular absence of screening of Coulomb interaction, which keeps its long-range form \cite{Screening,Review}. Various derivatives of graphene, such as twisted bilayer graphene \cite{bilayer}, are also characterized by Dirac dispersion, but in addition may have almost flat electronic band \cite{bilayer_disp}.

The effect of local Coulomb interaction in graphene was studied theoretically by variety of methods \cite{Herbut_2006,UInf1,UInf2,UInf3,UInf4} and was shown to produce spin-density wave (SDW) correlations, which at sufficiently strong interaction yield SDW long-range order. On contrary, the nearest-neighbor Coulomb repulsion favors formation of charge density wave (CDW) \cite{Herbut_2006,UInf1,TUfRG}. The effect of long-range Coulomb interaction was intensively investigated within the continuum model by mean-field \cite{Khveschenko,Gusynin,Murthy,Drut,Khveschenko_d,Gusynin_d,Gonzalez} and functional renormalization group (fRG) \cite{Katanin} approach and was also shown to yield CDW order at sufficiently strong interaction. 

At the same time, the most challenging theoretical problem is the description of the combined effect of short- and long-range interactions in graphene. It was argued recently that the screening of the non-local interaction in graphene by $\sigma$ bands essentially changes its physical properties, in particular shifts boundaries of spin and charge instabilities \cite{Ulybyshev_2013,OurFlakes}. This is related to the suppression of the non-local part of the Coulomb interaction, mainly between the nearest-neighboring and (to somewhat smaller degree) between the next nearest neighboring sites \cite{Wehling_2011}. Therefore, these interactions, together with the long-range tail, play important role in the physical properties of graphene.

To date, mostly methods treating finite clusters, such as quantum Monte-Carlo (QMC)   
\cite{Tang_2018,Ulybyshev_2013,Buividovich_2018,Strouthos,Smith_2014,Buividovich_2019} and fRG approach \cite{OurFlakes} were applied for treatment of simultaneous effect of local and non-local interactions in graphene. Among the methods, which consider infinite graphene sheet and treat the above mentioned interactions, the truncated unity fRG approach was applied \cite{TUfRG1}. The suppression of spin density wave instability by the non-local interaction, and the formation of CDW order for sufficiently large non-local interaction was investigated by the above mentioned approaches. Nevertheless, further development of the methods which allow to study the combined effect of local and non-local interactions in multi-sublattice systems is of great interest.

The dynamical mean-field theory (DMFT) \cite{DMFT_rev}, including the extended DMFT (E-DMFT) approach \cite{EDMFT_Si,EDMFT,EDMFT,EDMFT1}, as well as their non-local diagrammatic extensions \cite{OurRev}, provide powerful instruments for treatment of both, local and non-local interactions in correlated electronic systems. In the present paper we revisit the formulation of the E-DMFT approach for multi-sublattice systems \cite{EDMFT1}, and emphasize that it allows to treat explicitly both, the local and the non-local interaction inside the unit cell, which constitute the two most important components: on-site and nearest-neighbor interaction in case of graphene. The remaining part of the non-local interaction can be considered by introducing an effective retarded intra unit cell interaction of E-DMFT. Further treatment of the non-local correlations is possible in the spirit of the ladder non-local diagrammatic extensions of E-DMFT approach \cite{OurRev,MyEDMFT,DB,AbInitioDGA} by calculating the non-local susceptibility in the ladder approximation. For CDW instability this susceptibility also contains a correction for the difference between the actual non-local interaction and the effective retarded interaction of E-DMFT.   

We apply the above described approach to investigate the effect of non-local  Coulomb interaction in graphene and study magnetic and charge instabilities. The effects of both short- and long-range electron-electron interactions are studied. For calculations we use uniformly screened Coulomb interaction $\sim 1/(\epsilon r)$, where $r$ is the distance between electron positions. We also consider realistic non-local interaction, 
which has been determined by accurate first-principles calculations~\cite{Wehling_2011}. 
Considering, similarly to previous study in Ref. \cite{OurFlakes}, screening of the on-site and non-local components of interaction independently we obtain the phase diagram of graphene. 

Similarly to earlier results, we obtain the SDW (CDW) phases 
at sufficiently strong local (non-local) interactions. We show, that at a finite temperature the considered theory predicts the boundary of CDW instability, which is comparable to the result of fRG approach for finite clusters \cite{OurFlakes}. The obtained boundary of SDW instability agrees with QMC \cite{Tang_2018} and fRG \cite{OurFlakes} approaches, as well as the results of earlier approaches \cite{UInf1,UInf2,UInf3,UInf4} for purely local interaction. In general the tendency to SDW order is only slightly overestimated by the considered approach in comparison to previous approaches. In agreement with the previous study \cite{OurFlakes} 
the CDW instability is much stronger affected by the screening of Coulomb interaction by $\sigma$ bands, than the SDW instability, and the phase diagram shows the presence of a wide region with no instability.

Considering lower temperatures (which for finite systems would require larger lattice size) allows us to estimate the shift of phase boundaries with reducing temperature. The obtained shift of phase boundaries is found sufficiently small and does not change qualitatively previous results for the phase diagram and the conclusion of stability of semimetal (SM) state in freely suspended graphene. At the same time, we find pronounced spin density wave correlations in this state, which is not very far from the respective phase boundary.


The paper is organized as follows. In Sect. II, after presenting the model Hamiltonian we describe the considered E-DMFT method and evaluation of non-local susceptibilities. In Sect. III we consider the results for a purely local interaction, present phase diagram with account of non-local interaction, discuss stability of the CDW and SDW order and momentum dependence of the susceptibilities. 
Finally, Sect. IV  summarizes our main results and  {presents} conclusions. In Appendixes we present technical details of the solution of Bethe-Salpeter approach and calculation of Fourier transform of Coulomb interaction.
\section{Model and Method}
\begin{figure}[b]
		\center{\includegraphics[width=0.8\linewidth]{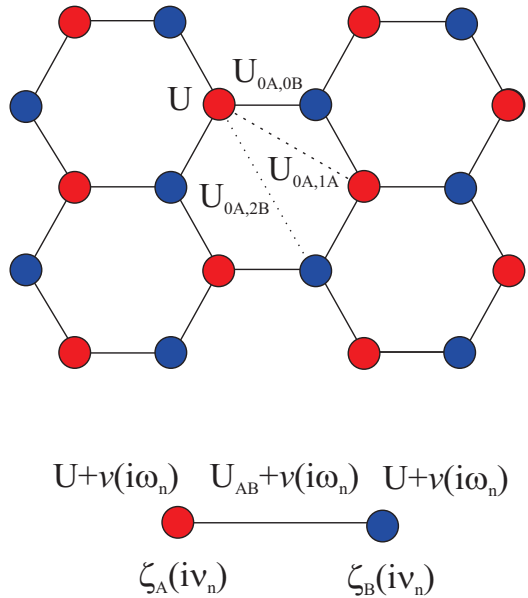}}
		\caption{{(Color online).} Upper part: Fragment of the hexagonal lattice with the on-site ($U$), nearest neighbor ($U_{0A,0B}$, solid lines), next  ($U_{0A,1A}$) and next to next ($U_{0A,2B}$) nearest neighbor  interactions (dashed lines) between A (red) and B (blue) sublattice sites. Lower part: the impurity model formed by $A$ and $B$ sublattice sites, and including the bath Green functions $\zeta_{A,B}(\iu \nu_n)$, acting at each site, the on-site interaction $U$, the intersublattice interaction $U_{AB}$ (shown by solid line), and the dynamic interaction $v(i\omega_n)$.}
\label{SystemsPic}
\end{figure}
We consider the two sublattice system, in particular, on the hexagonal lattice, with the hopping between sublattices and long-range interaction (see Fig. \ref{SystemsPic}). The corresponding Hamiltonian can be written as
\begin{align}
 \mathcal{H}&=-t\sum_{<ij>,\sigma}\left({\hat d}^{\dagger}_{iA\sigma}{\hat d}_{jB\sigma}+H.c.\right)\notag\\&+\dfrac{1}{2}\sum_{im,jm'}U_{im,jm'}\left({\hat n}_{im}-1\right)\left({\hat n}_{jm'}-1\right).
 \label{GNFHamiltonian}
\end{align}
Here, ${\hat d}^{\dagger}_{im\sigma}$ $({\hat d}_{im\sigma})$ is a creation (annihilation) operator of an electron at the unit cell $i$ and sublattice $m=A,B$ with a spin index  $\sigma=\uparrow,\downarrow$, ${\hat n}_{jm\sigma}={\hat d}^{\dagger}_{jm\sigma}{\hat d}_{jm\sigma}$ and ${\hat n}_{jm}={\hat n}_{jm\uparrow}+{\hat n}_{jm\downarrow}$, the summation in the first term of Eq.~(\ref{GNFHamiltonian}) is taken over nearest neighbor sites. 
 The last term in Eq.~(\ref{GNFHamiltonian}) describes the electron-electron interaction with the potential $U_{im,jm'}$  
 that includes both the on-site $U=U_{im,im}$ and non-local $U_{(i,m)\ne (j,m')}$ contributions.



For purely local interaction ($U_{(i,m)\ne (j,m')}=0$) the dynamical mean-field theory is formulated straightforwardly. For generity, we consider the possibility of spin density wave phase with opposite staggered magnetization at $A$ and $B$ sublattice. By mapping the local self-energy $\Sigma_{B,-\sigma}=\Sigma_{A,\sigma}\equiv\Sigma_\sigma$ we obtain the self-consistency condition for the local Green's function at the $A$ sublattice at half filling (see, e.g., Ref. \cite{DMFT_rev}):
\begin{equation}
G^{\rm loc}_{A\sigma}(i \nu_n)=\int\limits_0^{W} \frac{{\mathcal M}_{-\sigma}(\iu \nu_n)}{{\mathcal M}_{\uparrow}(\iu \nu_n){\mathcal M}_{\downarrow}(\iu \nu_n)-\epsilon^2}\rho(\epsilon)d\epsilon,  \label{SC_Ferm} 
\end{equation}
where ${\mathcal M}_{\sigma}(\iu \nu_n)=\iu\nu_n-\Sigma_{\sigma}(\iu \nu_n)+h\sigma/2$, $\rho(\epsilon)$ is the density of states, small magnetic field $h$ is introduced to obtain the spin symmetry breaking solution, $W$ is the upper band edge ($W=3t$ for the hexagonal lattice), $\nu_n$ are the fermionic Matsubara frequencies, $\iu$ is a complex unity. The position of the chemical potential at the particle-hole symmetric point is assumed to be at zero energy.



To take into account the non-local interaction,
we generalize the dynamical mean field theory in the following way. We first average the intra- and intersublattice interaction over momentum 
\begin{equation}
U_{mm'}=\sum_{\bf q} V_{mm'}({\bf q}),\label{Ust1}
\end{equation}
by using the respective Fourier transformed components 
\begin{equation}
V_{mm'}({\bf q})=\sum_{j} U_{im,jm'}e^{\iu {\bf q} {\bf r}_{im,jm'}},\label{Vmm1}
\end{equation}
${\bf r}_{im,jm'}$ is the radius-vector connecting corresponding lattice sites $i,m$ and $j,m'$. The average $U_{AA}$ and $U_{BB}$ interactions are equal to the corresponding local part $U$, and the only part, originating from the non-local interaction, is the average inter-sublattice interaction $U_{AB}$.

On the next step we account for the remaining non-local interaction ${\widetilde V}_{mm'}({\bf q})=V_{mm'}({\bf q})-U_{mm'}$ analogously to E-DMFT approach \cite{EDMFT_Si,EDMFT,EDMFT1}. Picking out the average interaction ensures us that $\sum_{\mathbf q} \widetilde{V}_{mm'}({\mathbf q})=0$. The action of the local model,
containing intra- ($U$) and inter ($U_{AB}$) sublattice static interaction and the effective dynamic interaction, which originates from the non-local interaction ${\widetilde V}({\bf q})$, reads (cf. Ref. \cite{EDMFT1} and Fig. \ref{SystemsPic}) 
\begin{align}
S_{\rm DMFT}&=-\sum_{m,\iu \nu_n}  \zeta_{m \sigma}(\iu \nu_n) d^{\dagger}_{im\sigma}(\iu \nu_n)d_{im\sigma}(\iu \nu_n)\notag\\
&+\frac{1}{2}\sum_{m\sigma,m'\sigma',\iu\omega_n}U^{\sigma,\sigma'}_{m m'}  n_{im\sigma}(\iu \omega_n)n_{im'\sigma'}(-\iu \omega_n)\notag\\
&+\frac{1}{2}\sum_{\iu\omega_n}v(\iu\omega_n)  n_{i}(\iu \omega_n) n_{i}(-\iu \omega_n),
\label{Simp}
\end{align}
where ${ d}^{\dagger}_{im\sigma}$ $({d}_{im\sigma})$ are Grassmann variables, $i$ refers to the impurity site,  $U_{mm}^{\sigma,\sigma}=0$, $U_{mm}^{\sigma,-\sigma}=U$, $U_{AB}^{\sigma,\sigma'}=U_{BA}^{\sigma,\sigma'}=U_{AB}$. The bath Green function $\zeta_{m\sigma}(\iu \nu_n)=[G^{\rm loc}_{m\sigma}(\iu \nu_n)]^{-1}+\Sigma_{m \sigma}(\iu \nu_n)$ 
is determined by the self-consistent condition (\ref{SC_Ferm}) (the self energy is spin- and sublattice independent in the spin symmetric phase and the mapping $m=B\rightarrow m=A$, $-\sigma\rightarrow\sigma$ in the SDW state is used), $n_{im\sigma}(\iu \omega_n)=\sum_{\nu_n} { d}^{\dagger}_{im\sigma}(\iu \nu_n) {d}_{im\sigma} (\iu \nu_n+\omega_n)$, $n_{i}(\iu \omega_n)=\sum_{m,\sigma} n_{im\sigma}(i \omega_n)$. 

The retarded interaction $v(\iu \omega_n)$ originates from the interaction ${\widetilde V}({\bf q})$ of the lattice model. The corresponding self-consistent equation for $v(\iu \omega_n)$ can be derived in a standard way (cf. Refs. \cite{EDMFT,EDMFT1}). We introduce the polarization operator matrix in the sublattice space
\begin{equation}
    \Pi(\iu\omega_n)=\left(
    \begin{array}{cc}
\Pi_{AA}(\iu \omega_n) & \Pi_{AB}(\iu \omega_n) \\ 
\Pi_{AB}(\iu \omega_n) & \Pi_{AA}(\iu \omega_n)%
\end{array}
    \right),
\end{equation} 
which is related
to the matrix local charge susceptibility  $\chi_{mm'}(\iu \omega_n)=(1/2)\sum_{\sigma\sigma'}\langle n_{im\sigma}(\iu \omega_n) n_{im'\sigma'}(-\iu \omega_n)\rangle$ by
\begin{equation}
    \chi(\iu \omega_n)=
    \left[\Pi(\iu \omega_n)^{-1}+2v_M(\iu \omega_n))\right]^{-1},
\end{equation}
where
\begin{equation}
    v_M(\iu\omega_n)=v(\iu \omega_n)\left(
    \begin{array}{cc}
1 & 1 \\ 
1 & 1%
\end{array}
    \right)
\end{equation}
(we assume the retarded interaction to be the same between the same and different sublattices). The self-consistent equation for $v(\iu \omega_n)$ reads
\begin{equation}
    S=\sum_{{\bf q}}{\rm Tr} \left\{
    {\widetilde V}({\bf q})
    \left[\Pi(\iu \omega_n)^{-1}+2{\widetilde V}({\bf q})\right]^{-1}
    \right\},
\end{equation}
where we consider $\widetilde{V}({\bf q})$ also as a matrix with respect to the sublattice indexes and 
\begin{align}
    S&=
    {\rm Tr} \left[
    v_M(\iu \omega_n) \chi(\iu \omega_n)
    \right]\notag\\
    &=\frac{1}{2}v(\iu \omega_n)\left[({\rm Tr} \Pi(\iu \omega_n))^{-1}+2 v(\iu \omega_n)\right]^{-1}.
\end{align} 
Picking the average interaction in Eqs. (\ref{Ust1}) and (\ref{Vmm1}) ensures that $v(\iu \omega_n)\rightarrow 0$ at $\omega_n\rightarrow \infty$.

To find spin- and charge instabilities we calculate the non-local charge (spin) susceptibility (cf. Refs. \cite{DB,MyEDMFT,AbInitioDGA,OurRev})
\begin{align}
\mathcal{\chi}_{q}^{c(s),mn,m'n'}&=\sum_{\nu,\nu'}\left[  (\chi_{q,\nu}^{0,mn,m'n'})_{mn,m'n'}^{-1} \delta_{\nu\nu^{\prime}}\right.\label{chi_BS}\\
&-\left.
\Phi_{q,\nu\nu^{\prime}}
^{c(s),mn,m'n'}+{V}_q^{c(s),m n,m'n'}\right]  _{\nu m n,
\nu^{\prime} m' n'}^{-1},\notag%
\end{align}
where ${V}_q^{c,mn,m'n'}=2(\widetilde{V}_{mm'}({\bf q})-v(\iu \omega_n))\delta_{mn}\delta_{m'n'}$, $V_q^{s}=0$, the bare susceptibility
\begin{equation}
\chi^{0,mn,m'n'}_{ q,\nu}=-T\sum\limits_{\mathbf{k}}G^{nm'}_{k}G^{n'm}_{k+q}\label{chi0}
\end{equation}
is considered as a matrix with respect to composite indexes $m,n$ and $m',n'$, $q=({\bf q},\iu \omega_n)$. The vertex $\Phi$ is evaluated from the local vertex $\Gamma$ via the Bethe-Salpeter equation
\begin{align}
{\Gamma}_{\omega,\nu \nu'}^{c(s),mn,m'n'}&=\left[  (\Phi_{\omega,\nu \nu' }^{0,mn,m'n'})_{\nu mn,\nu' m'n'}^{-1} \right.\label{local_BS}\\
&-\left.
 \chi_{\omega,\nu}^{0,mn,m'n'} \delta_{\nu\nu^{\prime}}\right]  _{\nu m n,
\nu^{\prime} m' n'}^{-1},\notag%
\end{align}
where $\chi_{\omega,\nu}^{0,mn,m'n'}=-T G^{\rm loc}(\iu \nu) G^{\rm loc}(\iu \nu+\iu \omega) \delta_{mn'} \delta_{m'n}$. In the vertices $\Gamma^{c(s),mn,m'n'}_{\omega,\nu\nu'}$ and $\Phi^{c(s),mn,m'n'}_{\omega,\nu\nu'}$ the sublattice indexes $m,m'$ correspond to the incoming particles with frequencies $\nu,\nu'+\omega$, while the indexes $n,n'$ refer to the outgoing particles with frequencies $\nu+\omega,\nu'$.  The details on the numerical solution of Bethe-Salpeter equations are presented in Appendix A. The momentum dependent staggered charge- and spin susceptibilities are then evaluated as $\chi_q^{c(s),{\rm st}}=\sum_{m,n}(-1)^{m+n} \mathcal{\chi}_{q}^{c(s),mm,nn}$.

 For numerical analysis, aiming at the study of the physical properties of graphene, we consider the hexagonal lattice with the on-site interaction $U$ ($U=3.4t$ for freely suspended graphene) and uniformly screened Coulomb electron interaction $U_{(im)\neq (jm')}= e^2/(\epsilon_{\rm eff} \epsilon r_{im,jm'})$, where the effective dielectric permittivity $\epsilon_{\rm eff}=1.41$ describes the screening of the Coulomb interaction by $\sigma$-orbitals in graphene~\cite{Wehling_2011}. We also present results for the realistic screened Coulomb interaction $U_{(im)\ne (jm')}= U^{*}_{im,jm'}/\epsilon$, 
where $U^{*}_{im,jm'}$ is the realistic non-local potential in graphene, calculated by ab initio approach in Ref.~\cite{Wehling_2011}. The latter potential accounts for the screening of Coulomb interaction of nearest- and next-nearest neighbors, $U^*_{0A,0B}=5.5$~eV and $U^*_{0A,1A}=4.1$~eV. At distances larger than the distance between second-nearest-neighbor lattice sites $r_{im,jm'}>\sqrt{3}a$ ($a=0.142$ nm is graphene's lattice constant) the realistic potential is approximated by $U^*_{im,jm'}=1/(\epsilon_{\rm eff} r_{im,jm'})$, as in Ref.~\cite{Ulybyshev_2013}. The efficient way of calculating the lattice sums of long range interaction is discussed in Appendix B. Similarly to Ref. \cite{OurFlakes}, the independent variation of the parameters $U$ 
and $\epsilon$ allows us to study the interplay between the on-site and non-local parts of the interactions. 

Since the considered approach lacks the renormalization of the Fermi velocity by Coulomb interaction, we represent the uniformly screened interaction in the form $U_{(im)\neq (jm')}= 3\alpha ta/(2\epsilon_{\rm eff} \epsilon r_{im,jm'})$ where $\alpha=e^2/(\hbar v_F)$ is graphene's ``fine structure" constant, $v_F=3ta/(2\hbar)$ is the Fermi velocity. We choose  $\alpha=2.18$, which corresponds to
the experimentally observable Fermi velocity~\cite{Graphene} $v_F=10^8$~cm/sec, and use $t=1$ as a unit of energy. The interactions of the realistic screened potential are rescaled with respect to the uniformly screened interaction according to the results of the ab initio calculations ~\cite{Wehling_2011}.
\par
For the solution of the impurity problem with purely local interaction we apply the numerical renormalization group (NRG) approach \cite{NRG} at a small temperature $T=0.001t$ using TRIQS-NRG Lyublyana interface package \cite{TRIQS}, at finite temperature and/or finite non-local interaction we apply the continous-time quantum Monte-Carlo (CT-QMC) approach, realized in iQIST package \cite{iQIST}. 
The local vertices $\Gamma$ are also evaluated within the CT-QMC method for 120-160 fermionic frequencies (both positive and negative) and zero bosonic frequency $\omega=0$. For obtaining spin symmetry breaking solutions we introduce small magnetic field $h=0.001t$.

\section{Results}



\subsection{Purely local interaction
}
\begin{figure}[b]
	\center{\includegraphics[width=1.0\linewidth]{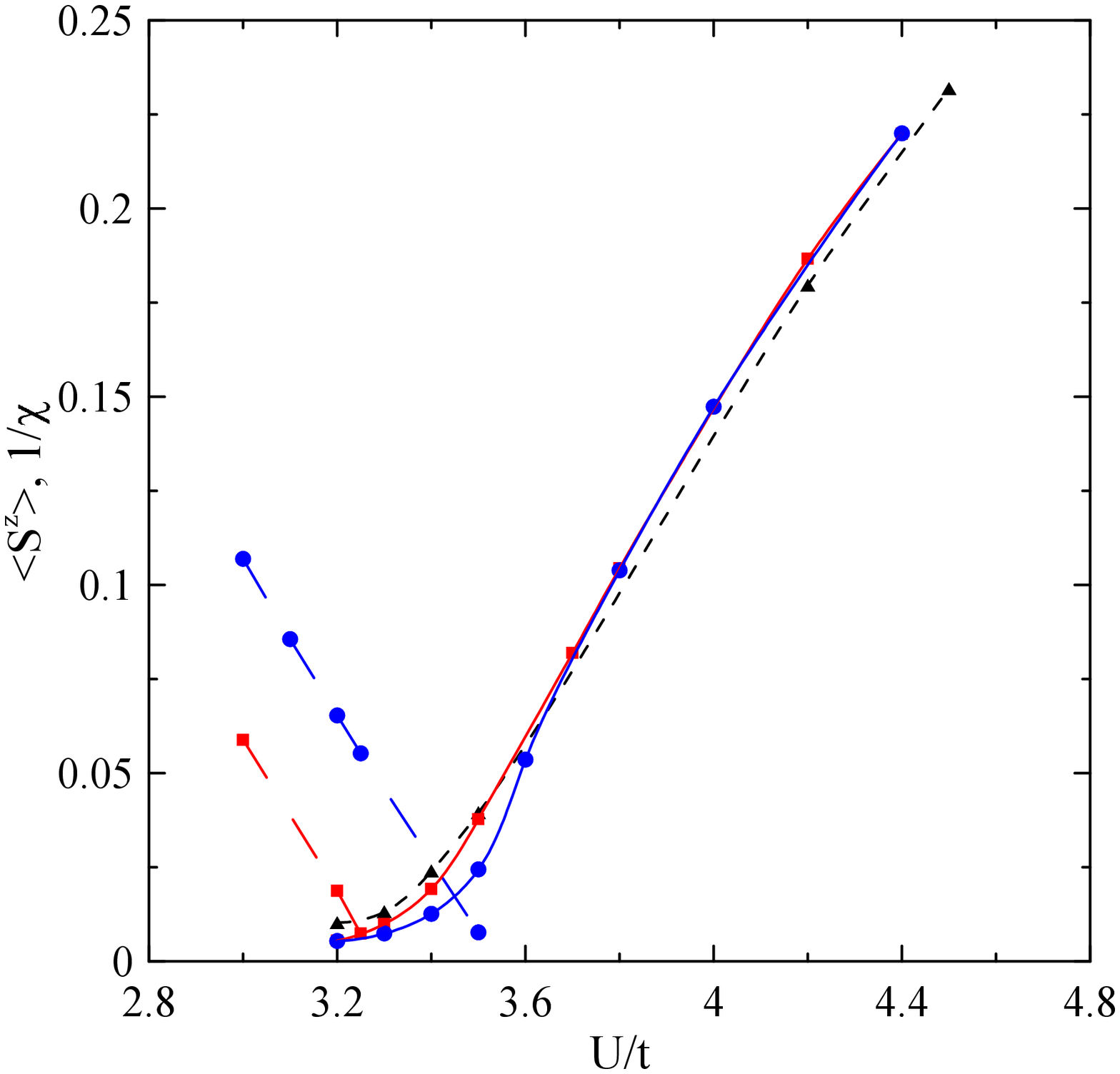}}
			\caption{(Color online) The average staggered magnetization $\langle S^z\rangle=|\langle\hat{n}_{im\uparrow}-\hat{n}_{im\downarrow}\rangle|/2$ as a function of $U/t$ for $T=0.05t$ (solid blue line with circles) and $T=0.02t$ (solid red line with squares), together with the inverse staggered susceptibilities $1/\chi^{s,{\rm st}}_0$ (respective long dashed lines) for purely local interaction $U_{(im)\ne (jm')}=0$. The dashed line with triangles shows the NRG result at $T=0.001t$. }
	\label{GNF54Local}
\end{figure}
Let us first consider the system with a purely local (on-site) interaction $U$, when $U_{(im)\ne (jm')}=0$. In Fig.~\ref{GNF54Local}, the results for the average staggered magnetization $\langle S^z \rangle=|\langle\hat{n}_{im\uparrow}-\hat{n}_{im\downarrow}\rangle|/2$ at $T=0.001t$, obtained within NRG, together with the sublattice magnetization and the staggered susceptibility at $T=0.02t$ and $T=0.05t$, calculated within CT-QMC approach (combined with the solution of Bethe-Salpeter equation for susceptibility) are shown. For small $U/t$ the magnetization is small but nonzero due to the presence of the finite magnetic field. 
With increasing $U/t$, the magnetization increases monotonously, indicating the formation of the SDW order. By comparing the results of CT-QMC and NRG approaches one can see that the position of phase transition at $T=0.02t$ is already sufficiently close to that at $T=0$. Because of the presence of finite magnetic field, however, accurate determination of the position of the phase transition from the magnetization only is difficult, since the phase transition is smoothed by magnetic field. At the same time, the calculation of staggered susceptibility is performed at zero magnetic field and allows us to determine the position of the phase transition. At $T=0.02t$ we obtain $U_c=3.3t$. 
This value is somewhat smaller than the result for the critical interaction $U_c=(3.7-3.8)t$, obtained in various approaches \cite{UInf1,UInf2,UInf3,UInf4}, which reflects mean-field nature of DMFT approach. Yet, it is in a reasonable agreement with the above cited results.

\begin{figure}[t]
		\center{\includegraphics[width=0.95\linewidth]{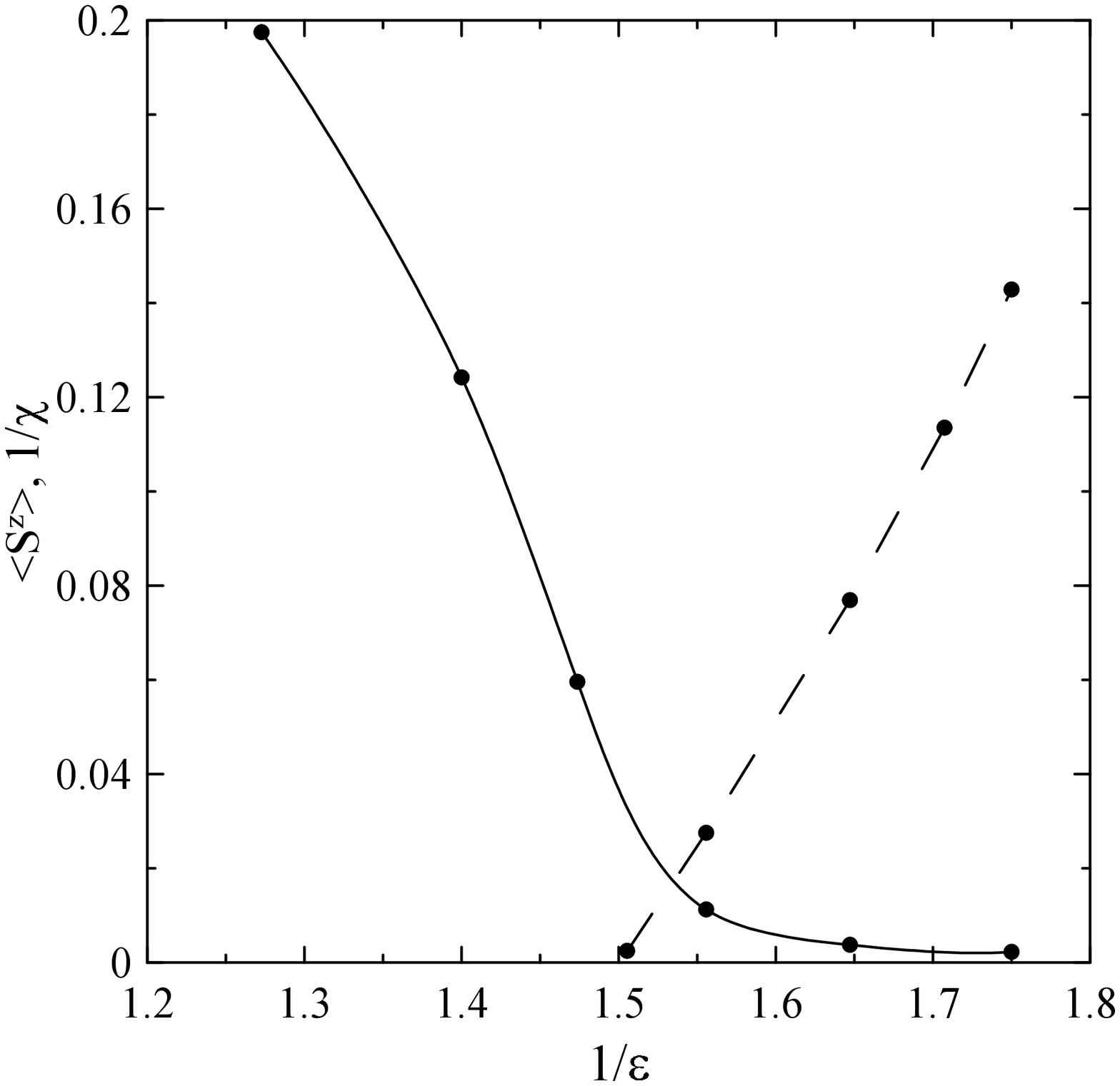}}
			\caption{The average staggered magnetization $\langle S^z\rangle$ (solid line) and the inverse staggered susceptibility $1/\chi^{s,{\rm st}}_0$ (dashed line) as functions of $1/\epsilon$ for $U=6t$, $T=0.05t$, and uniformly screened Coulomb interaction.}	\label{Sz1}
\end{figure}



\subsection{The effect of the non-local interaction
} 
We next study graphene with both {on-site} $U$ and non-local $U_{(im)\ne (jm')}$ electron-electron interactions, the latter is screened by the permittivity constant $\epsilon$.

In Fig.~\ref{Sz1}, we plot staggered magnetization $\langle S^z\rangle$ as a function of $1/\epsilon$ for $U=6t$, $T=0.05t$ and uniformly screened Coulomb interaction. With decrease of $\epsilon$ the sublattice magnetization is suppressed by the non-local interaction. One can see that similarly to the dependence of sublattice magnetization on $U/t$ in the absence of the non-local interaction, the phase transition is smoothed by the magnetic field, while calculation of the staggered  susceptibility allows to find the position of phase transition. 

\begin{figure}[t]
	\center{\includegraphics[width=1.0\linewidth]{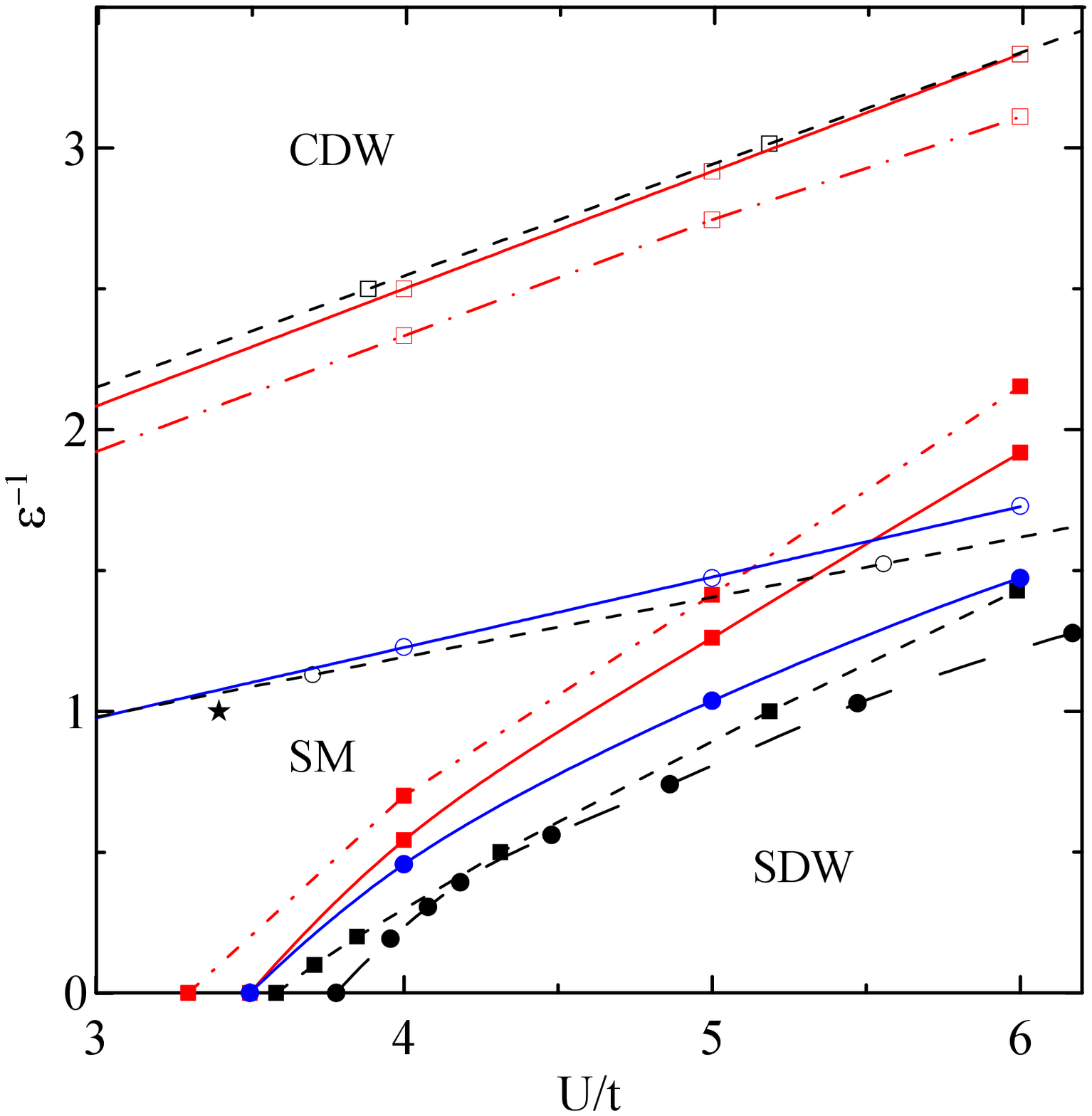}}
		\caption{(Color online) Phase diagram of the model (\ref{GNFHamiltonian}) in  $\left(U/t,\epsilon^{-1} \right)$ coordinates. SM-SDW phase transitions are denoted by lines with filled symbols, SM-CDW phase transitions are marked by lines with open symbols. Solid (dot dashed) lines represent the results of the approach of the present paper at $T=0.05t$ ($T=0.02t$), short dashed lines correspond to the results of the fRG approach of Ref. \cite{OurFlakes} for graphene nanoflakes with $N=96$ atoms, long dashed line is the scaling analysis of QMC results in Ref.~\cite{Tang_2018}. Lines marked by circles correspond to the uniformly screened $U_{i\neq j}=e^2/(\epsilon_{\rm eff}\epsilon r_{ij})$ form of the non-local interaction,
		the lines with squares correspond to the realistic screened $U_{i\neq j}= U^*_{ij}/\epsilon$ interaction.
		The point $\epsilon=1$ and $U=3.4t$, corresponding to freely suspended graphene, is marked by the star symbol.		}
			\label{PDs_AFM_PM}
	\label{PDs_main}
\end{figure}


Our results for the SM-SDW and SM-CDW phase-transition lines are summarized in Fig.~\ref{PDs_main}. For large $\epsilon$ the long-range part of the non-local potential is small and the on-site interaction $U$ plays the main role, such that SDW instability occurs at sufficiently large Coulomb repulsion $U$. When $\epsilon$ is sufficiently small, the SDW state becomes unstable and the SM state occurs. With decreasing $\epsilon$, the system undergoes the phase transition from the SM to the CDW ground state. 

For $T=0.05t$, the calculated SM-SDW and SM-CDW phase transition lines $\epsilon^{-1}(U/t)$  are close (especially for CDW instability) to those obtained previously within fRG approach for $N=96$ atoms graphene nanoflakes \cite{OurFlakes}. This is expectable, since finite temperature sets the cutoff of spin- and charge correlations, which acts similarly to the effect of finite size of the system; in fRG approach of Ref. \cite{OurFlakes} also the condition $\langle S^z \rangle_{T=0}=1/8$ was used to obtain positions of charge and spin instabilities, which mimics the effect of finite temperature. 
The obtained boundary of SDW instability for uniformly screened Coulomb interaction is also sufficiently close to that obtained from scaling analysis of QMC data for large systems ~\cite{Tang_2018,Note}. 

In agreement with the results of Ref. \cite{OurFlakes}, realistic screening of Coulomb interaction, having smaller nearest- and next-nearest neighbor interaction, only moderately increases critical inverse dielectric permittivity $1/\epsilon$ for SDW instability, but strongly enhances critical {non-local} interaction for the charge instability. As discussed in Ref. \cite{OurFlakes}, this provides  a wide region of the phase diagram with no instability. Graphene with the realistic non-local interaction ($U=3.4t$,  $\epsilon=1$) falls into this region and therefore corresponds to the SM ground state of the system (marked with the {star symbol} in Fig.~\ref{PDs_main}), which is far from both SM-SDW and SM-CDW phase-transition lines.

To study the effect of reducing temperature, we present for realistic interaction the phase boundaries at a lower temperature $T=0.02t$, for which is was shown in Fig. 1 that the results are expected to be close to the ground state properties. One can see that reducing temperature only moderately shifts boundaries of SDW and CDW regions towards the SM phase. At the same time, because of the mean-field nature of the considered E-DMFT approach we expect that this approach overestimates the tendency towards various instabilities. This especially holds for the SDW instability, since the non-local part of the interaction $\widetilde{V}({\mathbf q})$ is accounted in the position of this instability only via the ``average'' retarded interaction $v(\iu \omega_n)$. At the same time, for CDW instability $\widetilde{V}({\mathbf q})$ enters explicitly in the ladder summation in Eq. (\ref{chi_BS}), which provides better description of this instability.

\begin{figure}[b]
\center{\includegraphics[width=1.0\linewidth]{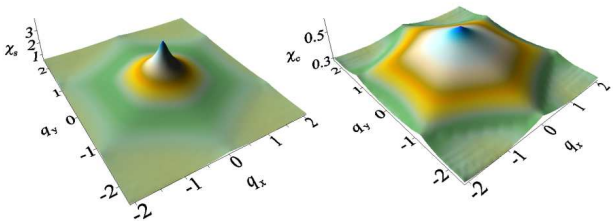}}
	\caption{(Color online) Momentum dependence of static staggered spin $\chi_{\bf q}^{s,{\rm st}}$ (left) and charge $\chi_{\bf q}^{c,{\rm st}}$ (right) susceptibility of freely suspended graphene ($U=3.4t$, $\epsilon=1$) at $T=0.02t$.}
	\label{Chiq}
\end{figure}

In Fig.~\ref{Chiq} we show the momentum dependence of charge- and spin staggered susceptibility $\chi^{c(s),{\rm st}}_{({\bf q},0)}$ for the realistic non-local potential ($U=3.4t$, $\epsilon=1$). The spin susceptibility shows a narrow peak with appreciable height $\sim 3/t$ near the momentum ${\bf q}=0$, corresponding to alternating magnetic field in $A$ and $B$ sublattice. This peak shows pronounced spin correlations, which occur due to presence of SDW instability not very far from the parameters of freely suspended graphene in the phase diagram of Fig. \ref{PDs_main}. The charge sublattice susceptibility shows broad peak with rather small height, which reflects large distance to the CDW instability in the phase diagram of Fig. \ref{PDs_main}.

\section{Conclusion}
In this paper, we have considered the formulation of extended dynamical mean-field theory for the two-sublattice systems, which allows us to study 
magnetic and charge instability of correlated systems. The main feature of the proposed approach, in comparison to general formulation of Ref. \cite{EDMFT1}, is picking out explicitly the average nearest neighbor (inter sublattice) interaction, whose importance for graphene was emphasized previously in Refs. \cite{Herbut_2006,UInf1,TUfRG}. The remaining part of the non-local interaction is considered by introducing an effective frequency-dependent intra-unit cell interaction, whose components between the same and different sublattices are considered to be the same. 
Based on the solution of E-DMFT problem, the local vertices were calculated and the non-local susceptibility was evaluated via the solution of the Bethe-Salpeter equation. For CDW instability, the corresponding susceptibility also accounts for the difference between the actual non-local interaction and an effective local frequency dependent one.

The developed method is applied to study SDW and CDW instabilities for varying on-site and non-local Coulomb interaction in graphene. The ground-state phase diagram at half-filling is obtained for the  uniformly screened, as well as the realistic screened long-range Coulomb interaction. The obtained phase diagram in $(U/t,\epsilon^{-1})$ coordinates, where parameter $\epsilon$ reduces the strength of the non-local interaction, is similar to the previously obtained within the fRG approach for finite graphene nanoflakes \cite{OurFlakes} and contains three phases: the semimetal (SM), the spin-density-wave (SDW), and the charge-density-wave (CDW) phase. The respective phase boundaries at the temperature $T=0.05t$ are found to be close to those obtained within the fRG approach of Ref. \cite{OurFlakes}. In particular, in agreement with previous study, 
the transition line between the SM and CDW phases is rather different for the realistic and uniformly screened long-range Coulomb interactions. We have therefore confirmed that the realistic screening of Coulomb interaction by $\sigma$ bands causes a strong enhancement of the critical value of long-range interaction needed to stabilize the CDW state. This results in a substantially wider region of stability of the SM phase for the case of the realistic non-local potential.  The obtained result on the SM-SDW transition line agrees with the scaling analysis of QMC data~\cite{Tang_2018}.

With reducing temperature to $T=0.02t$, we have found only slight shift of phase boundaries, which are expected to approach their ground state position in the considered approach. At the same time, this approach is expected to somewhat overestimate the tendency to various types of ordering. This especially concerns SDW instability, whose non-local susceptibility in the ladder approximation is not corrected by the difference of the non-local interaction and the effective frequency dependent local interaction, considered in E-DMFT. Nevertheless, the above mentioned comparison to the QMC data \cite{Tang_2018} shows that this overestimation is not particularly strong. As a future perspective, introducing the so called Moriyaesque lambda correction in the spin channel \cite{MyMoriya,OurRev} is possible to improve the account of spin fluctuations beyond E-DMFT. This, however, will require calculation at lower temperatures to obtain the position of quantum phase transition, since the finite temperature transition is suppressed in this case in agreement with the Mermin-Wagner theorem. Considering cluster extensions (see, e.g., Refs. \cite{Cluster1,Cluster2,Cluster3}) of the E-DMFT approach is another possible alternative of treatment of non-local correlations beyond E-DMFT.

Finally, evaluation of the non-local susceptibility for realistic interactions, corresponding to freely suspended graphene shows pronounced spin correlations, and somewhat weaker charge correlations.

The performed study shows that the proposed method can be used to describe magnetic and charge correlations in multi-sublattice systems. In this respect, its application to twisted bilayer graphene, as well as other derivatives of graphene, having Dirac form of the dispersion, is rather interesting. Evaluation of the non-local corrections to the self-energy (e.g. renormalization of the Fermi velocity, damping of electronic quasiparticles, etc.) within one of the approaches of Refs. \cite{DB,AbInitioDGA,MyEDMFT,OurRev} is also an interesting problem to study.

\section*{Acknowledgements} The author is grateful to V. Protsenko for stimulating discussions and careful reading of the manuscript. The author acknowledges the financial support from the Ministry of Science and Higher Education of the Russian Federation (Agreement No. 075-15-2021-606 and theme "Quant" AAAA-A18-118020190095-4). The work is also partly supported by RFBR grant 20-02-00252 A. The calculations were performed on the cluster of the Laboratory of material computer design of MIPT and the Uran supercomputer at the IMM UB RAS. 
\appendix
\section{Calculation of the non-local susceptibility}
To calculate the non-local susceptibility (\ref{chi_BS}) we use the method of Ref. \cite{My_BS}. To shorten the notations, we use the matrix form of the considered equations, written in the composite indexes of pair of sublattice indexes and frequencies $\mu=(\mathfrak{m},\nu)$, $\mu'=(\mathfrak{m}',\nu')$, where $\mathfrak{m}=(m_1,m_3)$ etc. is composite of two sublattice indexes.  In these notations, the equation for the triangular vertex $\Lambda_q^{c(s),\mu\mathfrak{m}'}=\langle d^+_{\mathbf{k}\mu} d_{\mathbf{k+q},\mu+\omega} n_{q,\mathfrak{m'}}^{c(s)}\rangle$ where $q=({\mathbf q},\iu \omega_n)$ and $n_{q,\mathfrak{m}}^{c(s)}=\sum_{k,\sigma,\sigma'}d^+_{k m_1 \sigma} \sigma^{0(z)}_{\sigma,\sigma'} d^+_{k+q,m_3 \sigma'} $ is the charge (z-component of spin) density ($\sigma^{0(z)}$ are the unity and $z$-component Pauli matrices, $d_{k m\sigma}$ is the Fourier transform of $d_{i m \sigma}$), reads
\begin{equation}
\Lambda _{q}^{c(s),\mu \mathfrak{m}'}=\sum\limits_{\nu ^{\prime }}\left[ I+\Gamma_q^{c(s)} \chi ^{0}_q%
\right] _{\mu \mu ^{\prime }},
\end{equation}
where $I$ is the identity matrix, the bare susceptibility $\chi^0_q$ is given by the equation (\ref{chi0}) of the main text, and assumed to be diagonal in the frequency indexes, the indexes $m_{1,2}$ and $m_{3,4}$ of the vertex $\Gamma_{q,\nu\nu'}^{m_1 m_3,m_2 m_4}$ (considered also as a matrix in fermionic frequencies and pairs of orbital indexes) correspond to the incoming and outgoing particles, respectively. At large frequency $\nu'\rightarrow \infty$ 
\begin{equation}
\Gamma_{q}^{c(s),\mu \mu'}\rightarrow (\Lambda _{q}^{c(s)}U_{q}^{c(s)})_{\mu \mathfrak{m}'},\label{GammaInf}
\end{equation}
(the right hand side is understood as a matrix product), where $U_{q}^{c}=-(U^{c}+2v^c(\iu \omega _{n}))$,  $U_{q}^{s}=U^s$, the non-zero components of interactions $U^{c(s)}$ are given by 
\begin{eqnarray}
U^{c(s),mm,mm}&=&U_{AA},\\
U^{c,mm,nn}&=&2U_{AB}~~(m\neq n),\notag\\
U^{c(s),mn,nm}&=&-U_{AB}~~(m\neq n),\notag\\
v^{c,m_1 m_3,m_2 m_4}(\iu \omega _{n})&=&v(\iu \omega _{n})\delta_{m_1 m_3}\delta_{m_2 m_4}.\notag
\end{eqnarray}
For  $\nu\rightarrow \infty$ the equation, transposed to the Eq. (\ref{GammaInf}) holds. In Eq. (\ref{GammaInf}) we have neglected the subleading terms, whose contribution was shown small in Ref. \cite{My_BS}.  Splitting frequency summation to the frequencies inside a chosen frequency box $B$ and outside of it, we find (cf. Ref. \cite{My_BS})
\begin{align}
&\Lambda _{q}^{c(s)}=\left[ I+\Gamma _{q}^{c(s)}\chi _{0}\right] \left[
E-U_{q}^{c(s)}X_{q}\right] ^{-1},\\
&\Lambda _{q}^{c(s),(\mathfrak{m},\nu\rightarrow \infty),\mathfrak{m}'}= \left[E+U_q^{c(s)}\chi_q^{c(s)} \right]_{\mathfrak{m}\mathfrak{m}'},
\end{align}
where $X_{q}^{\mathfrak{m}\mathfrak{m}'}=\sum_{\nu _{n}\notin B}\chi _{q,\nu}^{0,\mathfrak{m}\mathfrak{m}'}$, $E$ is the identity matrix with respect to pairs of orbital indexes, which does not depend on frequencies. Introducing the non-local susceptibility  $\overline{\chi}_q^{c(s),\mathfrak{m}\mathfrak{m}'} =\sum_{\nu}(\chi_q^{0}\Lambda_q^{c(s)} )_{\mu \mathfrak{m} ^{\prime }}$ and splitting again the frequency summation we find
\begin{align}
&\overline{\chi}_q^{c(s)} =\left[ E-X_q U_q ^{c(s)}\right] ^{-1}\left[
\chi_q ^{0}\Lambda_q^{c(s)} +X_q\right],\notag\\
E+U_q^{c(s)}&\overline{\chi}_q^{c(s)} =\left[ E-U_q^{c(s)}X_q%
\right]^{-1}\left[ E+U_q ^{c(s)}\chi_q ^{0}\Lambda_q^{c(s)} \right].
\end{align}
For the reduced triangular vertex $\gamma_q^{c(s)} ={\Lambda_q^{c(s)}\left[ E+U_q^{c(s)}\overline{\chi}_q^{c(s)} \right] ^{-1}}$ we therefore find after algebraic manipulations
\begin{equation}
    \gamma_q^{c(s),\mu \mathfrak{m}'}=\sum\limits_{\nu' \in B} \left\{\left[ I+\Lambda_q^{c(s)} U_q^{c(s)}\chi ^{0}_q\right] ^{-1}\left[ I+\Gamma_q^{c(s)} \chi ^{0}_q\right]\right\}_{\mu \mu'} 
\end{equation}
Using the Bethe-Salpeter equation (\ref{local_BS}) of the main text, restricted to the considered frequency box (which serves as the definition of the irreducible vertex $\Phi$, cf. Ref. \cite{My_BS}), we find
\begin{equation}
    \gamma_q^{c(s),\mu \mathfrak{m}'}=\sum\limits_{\nu' \in B} \left[I-\Phi_q^{c(s)} \chi ^{0}_q+\widetilde{U}_q^{c(s)} \chi ^{0}_q\right]_{\mu \mu'} ^{-1},\label{gamma}
\end{equation}
where $\widetilde{U}_q^{c(s)}=[E-U_q^{c(s)} X_q^{c(s)}]^{-1}U _q^{c(s)}$.

Performing further the same steps as in Ref. \cite{My_BS} we find that the irreducible susceptibility $\phi^{c(s)}$ fulfilling the matrix relation
\begin{equation}
\overline{\chi}^{c(s)}_q=\left[E-\phi_q^{c(s)}U_q^{c(s)}\right]^{-1}\phi_q^{c(s)} \label{phi}
\end{equation}
is given by
\begin{eqnarray}
\phi_q^{c(s)}  &=&\left[ E+\chi_q ^{0}\Gamma_q^{c(s)} U_q^{c(s)}\right] ^{-1}\left[ \chi_q ^{0}\Gamma_q^{c(s)} +X_q%
\right]   \nonumber \\
&=&\chi_q ^{0}\gamma_q^{c(s)} +X_q. \label{phi1}
\end{eqnarray}
Equation (\ref{gamma}) and second line of Eq. (\ref{phi1}) are used in numerical calculations of the irreducible susceptibility $\phi_q^{c(s)}$. In the spirit of the non-local extensions of dynamical mean-field theory \cite{OurRev,DB,AbInitioDGA,MyEDMFT} we calculate the non-local susceptibility by substituting in Eq. (\ref{phi}) $U_{q}^{c}=-(U^{c}+2\widetilde{V}^c({\mathbf q}))$, where $\widetilde{V}^{c,m_1 m_3,m_2 m_4}({\mathbf q}))=\widetilde{V}_{m_1 m_2}({\mathbf q})\delta_{m_1 m_3}\delta_{m_2 m_4}$, and  $U_{q}^{s}=U^s$. 

\section{Calculation of lattice sums}
To calculate the lattice sum%
\begin{equation}
V_{AA}^{\text{C}}(\mathbf{q})=\sum\limits_{j\in A}{}^{\prime }\frac{e^{\text{%
i}\mathbf{qR}_{j}}}{R_{j}},
\end{equation}%
where the site index $j$ runs over the sites of $A$ sublattice (the dash stands for the excluding ${\bf R}_j=0$ term from the sum), we use the method of Ref. \cite{LS}. In
particular, we introduce the functions 
\begin{eqnarray}
w(\mathbf{q,r}) &=&\sum\limits_{j\in A}{}^{\prime }\delta (\mathbf{r-R}%
_{j})e^{\text{i}\mathbf{qR}_{j}}  \nonumber \\
&=&\overline{w}(\mathbf{q},\mathbf{r})-\delta (\mathbf{r}), \\
\overline{w}(\mathbf{q},\mathbf{r}) &=&\sum\limits_{j\in A}{}\delta (\mathbf{%
r-R}_{j})e^{\text{i}\mathbf{qR}_{j}},
\end{eqnarray}%
and rewrite $V_{AA}^{\text{C}}(\mathbf{q})$ as%
\begin{eqnarray}
V_{AA}^{\text{C}}(\mathbf{q}) &=&\frac{1}{\Gamma (1/2)}\left[ \int w(\mathbf{%
q,r})\frac{\Gamma (1/2,\pi r^{2})}{r}d^{2}\mathbf{r}\right.   \nonumber \\
&&\left. +\int w(\mathbf{q,r})\frac{\gamma (1/2,\pi r^{2})}{r}d^{2}\mathbf{%
r}\right]  \\
&=&\frac{1}{\Gamma (1/2)}\left[ \sum\limits_{j\in A}{}^{\prime }\frac{\Gamma
(1/2,\pi R_{j}^{2})}{R_j}\right.   \nonumber \\
&&\left. +\int \overline{w}(\mathbf{q,r})\frac{\gamma (1/2,\pi r^{2})}{r}%
d^{2}\mathbf{r-}2\sqrt{\pi }\right], 
\end{eqnarray}%
where $\Gamma (n,x)$ is the incomplete gamma-function and $\gamma
(n,x)=\Gamma (n)-\Gamma (n,x).$ Applying Parseval's formula 
\begin{equation}
\int \overline{w}(\mathbf{q,r})e^{\iu\mathbf{hr}}d^{2}\mathbf{r=}\frac{(2\pi
)^{2}}{v_{a}}\sum\limits_{\lambda }\delta (\mathbf{h}+\mathbf{q}-\mathbf{h}%
_{\lambda }),
\end{equation}%
where $\mathbf{h}_{\lambda }$ are the reciprocal lattice vectors, $v_a$ is the volume of the unit cell, we find%
\begin{eqnarray}
V_{AA}^{\text{C}}(\mathbf{q}) &=&=\frac{1}{\Gamma (1/2)}\left[
\sum\limits_{j\in A}{}^{\prime }\frac{\Gamma (1/2,\pi R_{j}^{2})}{R_j}%
\right.    \\
&&\left. +\frac{1}{v_{a}}\int \sum\limits_{\lambda }\delta (\mathbf{h}+%
\mathbf{q}-\mathbf{h}_{\lambda })F(h)d^{2}\mathbf{h-}2\sqrt{\pi }\right], \nonumber
\end{eqnarray}%
where%
\begin{eqnarray}
F(h) &=&\int \frac{\gamma (1/2,\pi r^{2})}{r}e^{\iu\mathbf{hr}}d^{2}\mathbf{r%
}  \nonumber \\
&=&2\pi \Gamma (1/2,h^{2}/4\pi )/h.
\end{eqnarray}%
Performing integration we find%
\begin{eqnarray}
V_{AA}^{\text{C}}(\mathbf{q}) &=&\frac{1}{\Gamma (1/2)}\left[
\sum\limits_{j\in A}{}^{\prime }\frac{\Gamma (1/2,\pi R_{j}^{2})}{R_j}%
\right.   \\
&&\left. +\frac{2\pi }{v_{a}}\sum\limits_{\lambda }\frac{\Gamma (1/2,|%
\mathbf{h}_{\lambda }-\mathbf{q}|^{2}/4\pi )}{|\mathbf{h}_{\lambda }-\mathbf{%
q}|}\mathbf{-}2\sqrt{\pi }\right].  \nonumber
\end{eqnarray}%
Similarly, we find%
\begin{eqnarray}
V_{AB}^{\text{C}}(\mathbf{q}) &=&\sum\limits_{j\in A}{}\frac{e^{\text{i}%
\mathbf{q(R}_{j}+\mbox
{\boldmath $\delta $})}}{|\mathbf{R}_{j}+\mbox
{\boldmath $\delta $}|} \\
&=&\frac{e^{\text{i}\mathbf{q}\mbox
{\boldmath $\delta $}}}{\Gamma (1/2)}\left[ \int w(\mathbf{q,r})\frac{\Gamma
(1/2,\pi (\mathbf{r}+\mbox
{\boldmath $\delta $})^{2})}{|\mathbf{r+}\mbox
{\boldmath $\delta $}|}d^{2}\mathbf{r}\right.  \notag \\
&&\left. +\int w(\mathbf{q,r})\frac{\gamma (1/2,\pi (\mathbf{r}+%
\mbox
{\boldmath $\delta $})^{2})}{|\mathbf{r+}\mbox
{\boldmath $\delta $}|}d^{2}\mathbf{r}\right]  \notag \\
&=&\frac{1}{\Gamma (1/2)}\left[ \sum\limits_{j\in A}{}e^{\text{i}\mathbf{q(R}%
_{j}+\mbox
{\boldmath $\delta $})}\frac{\Gamma (1/2,\pi (\mathbf{R}_{j}+%
\mbox
{\boldmath $\delta $})^{2})}{|\mathbf{R}_{j}\mathbf{+}%
\mbox
{\boldmath $\delta $}|}\right.  \notag \\
&&\left. +\frac{1}{v_{a}}\int \sum\limits_{\lambda }\delta (\mathbf{h}+%
\mathbf{q}-\mathbf{h}_{\lambda })F(h)e^{\iu(\mathbf{h+q)}%
\mbox
{\boldmath $\delta $}}d^{2}\mathbf{h}\right],  \notag \\
 &=&\frac{1}{\Gamma (1/2)}\left[ \sum\limits_{j\in A}{}e^{\text{i}\mathbf{q(R}%
_{j}+\mbox
{\boldmath $\delta $})}\frac{\Gamma (1/2,\pi (\mathbf{R}_{j}+%
\mbox
{\boldmath $\delta $})^{2})}{|\mathbf{R}_{j}\mathbf{+}%
\mbox
{\boldmath $\delta $}|}\right. \notag \\
&&\left. +\frac{2\pi }{v_{a}}\sum\limits_{\lambda }\frac{\Gamma (1/2,|%
\mathbf{h}_{\lambda }-\mathbf{q}|^{2}/4\pi )}{|\mathbf{h}_{\lambda }-\mathbf{%
q}|}e^{\iu\mathbf{h}_{\lambda }\mbox
{\boldmath $\delta $}}\right]. \notag
\end{eqnarray}%
where the vector $\mbox{\boldmath $\delta$}$ connects $A$ and $B$ site within the unit cell. The resulting components of the Fourier-transformed interaction for the bare
Coulomb potential read%
\begin{eqnarray}
V_{AA}(\mathbf{q}) &=&U+V_{AA}^{\text{C}}(\mathbf{q})/(\epsilon _{\mathrm{eff%
}}\epsilon )  \label{VAB} \\
V_{AB}(\mathbf{q}) &=&V_{AB}^{\text{C}}(\mathbf{q})/(\epsilon _{\mathrm{eff}%
}\epsilon )
\end{eqnarray}%
In case of realistic Coulomb potential we substract from Eqs. (\ref{VAB})
the contribution of nearest- and next-nearest neighbors and add instead
their realistic values, $U_{0A,0B}e^{\text{i}\mathbf{q(R}_{j}+%
\mbox
{\boldmath $\delta $})}/\epsilon $ and $U_{0A,1A}e^{\text{i}\mathbf{qR}%
_{j}}/\epsilon $.

\end{document}